\def\year{2018}\relax
\documentclass[letterpaper]{article} 
\usepackage{aaai18}  
\usepackage{times}  
\usepackage{helvet}  
\usepackage{courier}  
\usepackage{url}  
\usepackage{graphicx}  
\usepackage[usenames, dvipsnames]{color}
\usepackage{soul}
\usepackage{csquotes}

\begin{document}
%
\title{Mobilizing the Trump Train:\\Understanding Collective Action in a Political Trolling Community}


%
\author{Claudia Flores-Saviaga\textsuperscript{1}, {Brian C. Keegan}\textsuperscript{2}, {Saiph Savage}\textsuperscript{1,3}\\
\textsuperscript{1}West Virginia University\\
\textsuperscript{2}University of Colorado Boulder\\
\textsuperscript{3}Universidad Nacional Autonoma de Mexico (UNAM)\\ 
 {cif0001@mail.wvu.edu,  brian.keegan@colorado.edu, saiph.savage@mail.wvu.edu}}

\maketitle
\begin{abstract}
Political trolls initiate online discord not only for the lulz (laughs), but also for ideological reasons, such as promoting their desired political candidates. Political troll groups recently gained spotlight because they were considered central in helping Donald Trump win the 2016 US presidential election, which involved difficult mass  mobilizations. Political trolls face unique challenges as they must build their own communities while simultaneously disrupting others. However, little is known about how political trolls mobilize sufficient participation to suddenly become problems for others. We performed a quantitative longitudinal analysis of more than 16 million comments from one of the most popular and disruptive political trolling communities, the subreddit \texttt{/r/The\_Donald} (\texttt{T\_D}). We use \texttt{T\_D} as a lens to understand participation and collective action within these deviant spaces. In specific, we first study the characteristics of the most active participants to uncover what might drive their sustained participation. Next, we investigate how these active individuals mobilize their community to action.  Through our analysis we uncover that the most active employed distinct discursive strategies to mobilize participation, and deployed technical tools like bots to create a shared identity and sustain engagement. We conclude by providing data-backed design implications for designers of civic media. 
\end{abstract}

\section{Introduction}

The short history of politics and information technology suggests a pattern: there is a four year gap between the capabilities of information technology and their adoption in U.S. presidential campaigns. The web existed in 1996, but did not become relevant until the 2000 campaign; blogs existed in 2000; but were not influential until 2004; Facebook existed in 2004, but only made a splash in 2008; Twitter existed in 2008, but did not become essential until 2012. What then should we make of the 2016 presidential campaign and the 2012-era technologies they adopted? We argue that information filtering platforms like reddit, which demonstrated their capabilities for coordinating influential collective action before 2012, were politically weaponized in 2016 in the form of \textit{trolling communities} like \texttt{/r/The\_Donald} (\texttt{T\_D}). Trolling is hardly a new phenomenon~\cite{phillips2015we}. However, socio-technical systems like reddit, have technical affordances and popular influence that have enabled new forms of sustained disruption. However, we have yet to understand this new emerging complex organization in depth. 

While prior work has documented trolls' extreme tactics \cite{summit2016we} or explored methods for detecting and moderating trolls \cite{kumar2017antisocial}, online deviance researchers have overlooked how trolls must fulfill similar social and technical prerequisites to succeed like any other online community. We lack an understanding how these communities ``successfully'' self-organize to cause harm. We argue troll communities manage the same fundamental challenges faced by other online communities, especially sustaining participation and initiating collective action~\cite{kraut2012building}. The success of troll communities allows them to become problems for other online communities by undermining generalized norms of credibility, trust, and respect, which are essential for democratic self-governance. A deeper understanding of how political trolls mobilize will enable better preparations for when 2016 socio-technical capabilities are applied in 2020 elections.

In this paper we conduct a large scale empirical analysis on one of the most active and largest political trolling communities during the 2016 US presidential election: the subreddit \texttt{/r/The\_Donald} \cite{WikiTheDonald}. \texttt{T\_D} received significant media coverage due to its coordinated trolling efforts intended to disrupt and harass Trump's opponents     \cite{Howanarm49:online,Trollsfo10:online}. Participants of \texttt{T\_D} trolled in various ways: they organized Netflix boycotts after Netflix promoted TV shows opposing their political views \cite{BoycottN2:online,PeopleAr15:online}, orchestrated one-star Amazon reviews for Megyn Kelly's book \cite{ProTrump86:online,AmazonRe86:online}. Members of \texttt{T\_D} also organized the ``Great Meme War'' to harass Trump's detractors and flood the Internet with pro-Trump, anti-Hillary Clinton propaganda \cite{WorldWar4:online}. Participants of \texttt{T\_D} also actively promoted the use of satirical hashtags, such as  \#DraftOurDaughters to troll Hillary Clinton's initiative about supporting women to register for the military draft \cite{ClintonI67:online} or \#ShariaOurDaughters to take Islam ideologies to an absurd extreme \cite{NewMemeC14:online}.

We use \texttt{T\_D} as a lens to understand how a political trolling community manages its own challenges around sustaining participation and mobilization at scale. We conduct an empirical analysis on more than 16 million comments posted over almost two years on \texttt{T\_D} to understand these research questions: 

\begin{enumerate}
    \item What are the behavioral patterns of the most active participants in a political trolling community?
    \item How does a political trolling community mobilize its members to action?
   
 \end{enumerate}


Unlike other online communities, \texttt{T\_D} created socio-technical tools that integrated bots, gamified mechanisms, and ``calls to action'' to promote a shared identity and to motivate participation. The most effective calls to action provided a detailed understanding of why the community needed to participate. These findings have implications for developing more effective moderation strategies to govern trolling communities, identifying calls to action likely to lead to disruptive behavior, and designing more inclusive civic media technologies. 

\section{Background}
Our research draws from two areas of prior literature: (1) how social media facilitates collective action and (2) how deviant and trolling behaviors develop in online communities. This background identifies a reciprocal gap in our understanding of how collective action processes can be used towards deviant ends as well as how deviant online communities must fulfill the similar social and technical prerequisites to succeed like any other online community.

\subsection{Social Media Facilitating Collective Action.}
Collective action constitutes efforts done by a group of people to achieve a common goal~\cite{olson2009logic}. Social media facilitates collective action by offering: 

{\bf Mobilizing  Structures:} Social media influences people to take action in support of a cause, such was the case in the ``\#BlackLivesMatter'', ``Occupy'' or ``Arab Spring'' movements. Social media facilitated the mobilization of street protests by providing reliable communication channels as well as the social proof to encourage others to mobilize \cite{de2016social}. 

{\bf Opportunity Structures:} Social media promotes (or hinders) collective action by creating (or destroying) opportunities to communicate needs and to discover allies. Requests for help can unexpectedly be met with assistance, which enables positive feedback loops of reciprocal support that legitimizes the action \cite{althoff2014ask}.

{\bf Framing Processes:} Social media provides structures for making sense of the reality surrounding a collective effort. Being able to negotiate the interpretations and meanings of a collective effort is important because it provides a way to legitimate or motivate the actions of the group. For instance, some women have been using social media to interpret the street harassment they experience, to then create effective campaigns for fighting back against street harassment \cite{dimond2013hollaback}.

While social media appear to facilitate collective action in politics \cite{matias2016going}, less is known about the ability for deviant sub-communities to use leverage technological capabilities and coordinated social practices to support regressive, anti-social, or other disruptive online collective action. How are online communities like sub-reddits used by political trolls to produce collective action? What are their opportunity and mobilization structures? what do their framing processes look like?

\subsection{Deviance and Trolling in Online Communities}

Trolls operate within social contexts that must fulfill similar social and technical prerequisites to succeed like any other online community. Despite the dysfunction they visit upon others, deviant sub-communities within 4Chan~\cite{bernstein20114chan,hine2017kek}, Something Awful~\cite{pater_awful_2014}, or reddit~\cite{massanari2017gamergate} have their own norms, mechanisms for regulating user behavior, and socializing newcomers. While previous work has documented the extreme acts conducted by these trolling communities \cite{shachaf2010beyond}, our understanding of how these communities organize to cause harm is incomplete.

The targets of organized trolling efforts can experience a wide range of consequences: depression, helplessness, anxiety, low levels of self-esteem, frustration, insecurity and fear~\cite{cheng2017anyone,coles2016trolling}. More extreme cases of trolling behavior includes the case of \#GamerGate, where anonymous trolls collectively targeted female gamers, doxxed them, and made violent threats with the goal of driving them off social media platforms to silence their message~\cite{massanari2017gamergate}. Hacking, trolling, and other forms of cyber-disruption have also taken on geopolitical dimensions as national governments invest in covertly spreading or undermining messages. The Russian government has supported efforts to discredit leaders and activists opposed to the Putin government~\cite{SalutinP1:online}. Large-scale trolling campaigns supporting Donald Trump's candidacy for president coordinated raids on opponents' online communities. These political trolls strategically spread misinformation, and generated content to mock the opposition \cite{Trollthe6:online}. The ``Fifty Cent Army'' behind China's Great Firewall rigorously enforce trolling type messages against the limited dissent permitted~\cite{han2015defending}. 

The majority of research around online trolling communities has focused on characterizing their anti-social behavior~\cite{cheng2015antisocial}, identifying trolls~\cite{gardiner2016dark,kumar2017army,samory2017sizing}, or predicting users at risk of adopting trolling behavior~\cite{cheng2017anyone}. There remains a critical need to understand how trolls organize collective action and manage to carry out their disruptions that are sustained, focused, and unfortunately effective despite concerted moderation efforts. 

Qualitative research on political trolls has emphasized their motivations to participate and tactics to target their opponents  \cite{sanfilippo2017managing,bradshaw2017troops,coleman2014hacker,phillips2015we}. While some trolls are in it for the ``lulz'' \cite{coleman2014hacker} (i.e., to provoke Internet users for the laughs), political trolls are usually in it for ideological reasons. Tactics of political trolls include baiting ideological opponents into arguments or coordinating ``civic spamming'' campaigns \cite{Russiani87:online,1000Paid53:online}. 

We build off this prior work to now conduct a large scale quantitative analysis to understand how these deviant groups manage the same fundamental challenges of motivating and governing participation faced by traditional online communities in their own communities \cite{kraut2012building}.

\begin{figure}
\centering
  \includegraphics[width=1.0\columnwidth]{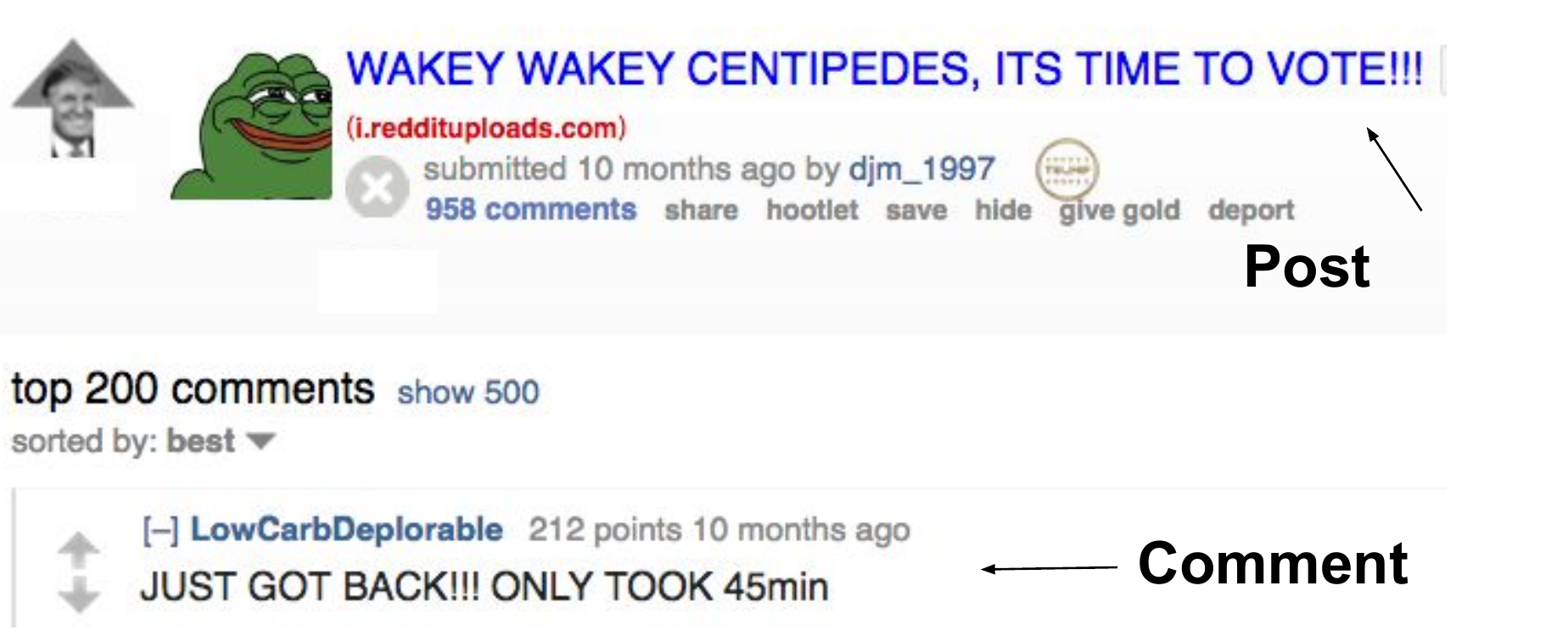}
  \caption{Example of reddit Post.}~\label{fig:redditpost}
\end{figure}

\section{Data Collection}
Reddit was created in 2005 by Steve Huffman and Alexis Ohanian as a community-driven platform for discussion, news aggregation and content rating~\cite{singer2014evolution}. The platform is composed of thousands of sub-communities (``subreddits'') focused on different topics. People on reddit can submit posts to a subreddit and others can up- or down-vote the posts as well as comment on them.  Figure~\ref{fig:redditpost} provides an example of a post with a comment.

We focus on the subreddit \texttt{/r/The\_Donald} (\texttt{T\_D})\footnote{\url{https://www.reddit.com/r/The_Donald/}}, which originated around the time Donald Trump announced his presidential run in June 2015. This subreddit followed all of Donald Trump's presidential run, his presidential win, and all of his actions and controversies in his presidential administration. The interface on \texttt{T\_D} is similar to an old Geocities website from the 1990s, complete with an animated GIF that has Donald Trump winking when you direct your mouse over him. \texttt{T\_D} had over 468,000 ``centipedes,'' or subscribers \footnote{\url{http://redditmetrics.com/r/The_Donald}}, and typically has around 5,000 visitors at any given time \cite{WikiTheDonald}.

\texttt{T\_D} generated controversial within the reddit community, across social media, and in political journalism during the 2016 presidential election. Many reddit users complained about the presence of \texttt{T\_D}, calling the subreddit's content ``hateful'' and claiming that it drowned out more substantive political deliberation. A substantial amount of content posted to \texttt{T\_D} allegedly originated from  ``alt-right'' users, who supported president Donald Trump, while participating in racist, sexist, Islamophobic, and other anti-social subreddits~\cite{lyons2017ctrl}. \texttt{T\_D} was labeled as one of the most active and largest political troll communities \cite{WikiTheDonald}.


We used a public dataset\footnote{\url{http://saviaga.com/the_donald-dataset/}} of all of the posts, comments, upvotes, downvotes of \texttt{T\_D} from June 30th, 2015 to February 28th 2017. We collected 16,349,287 comments from 342,731 participants. 

\section{RQ1: Behavioral Patterns of the Most Active}

Research Question 1 asked, ``what are the behavioral patterns of the most active participants in a political trolling community?'' We examined the most-active users by commenting activity in the \texttt{T\_D}. Commenting is a good indicator of involvement because it demands more engagement than simply up-voting/down-voting content. Understanding these individuals in greater depth becomes especially important as political troll communities depend on active contributions to carry out their disruptive mission. 

\subsection{Method.} 
We identified \texttt{T\_D} users who commented three times above the standard deviation of the average user's commenting behavior on the subreddit. We identified 3,427 individuals who generated  6,251,857 comments. Next, we analyzed the distinct words, slang, profanity, public figures and organizations used in each user's comments comments. We study these metrics given that prior work has identified they can serve to characterize behavioral patterns of subversive groups \cite{savage2015participatory}.

We identify distinct words using the TF-IDF metric and manually created a list of full, common, and nicknames and then flagged any posts or comments that mentioned any of the names to measure how users mentioned tokens related  to organizations or public figures involved in the 2016 elections. We also collected the names from Wikipedia \footnote{\url{http://bit.ly/2Hq1zmf}}, and news articles on Trump's nicknames for his opponents  \cite{Trumpsni79online,Presiden26online}.  
We went through each of the Wikipedia and news articles articles and added or merged the alternate names for each person.  For example, Trump called Senator Elizabeth Warren ``Pocahontas'' and ``Kim Jong-un'' was ``Little Rocket Man''. To identify slang we used a list of words known to represent the unique vocabulary of Trump supporters \cite{pede,pede2,pede3}. To identify profanity we used a public library that given a text can state its number of swear words\footnote{\url{http://bit.ly/2qkZ1OU}}. 

\begin{figure}
\centering
  \includegraphics[width=1.0\columnwidth]{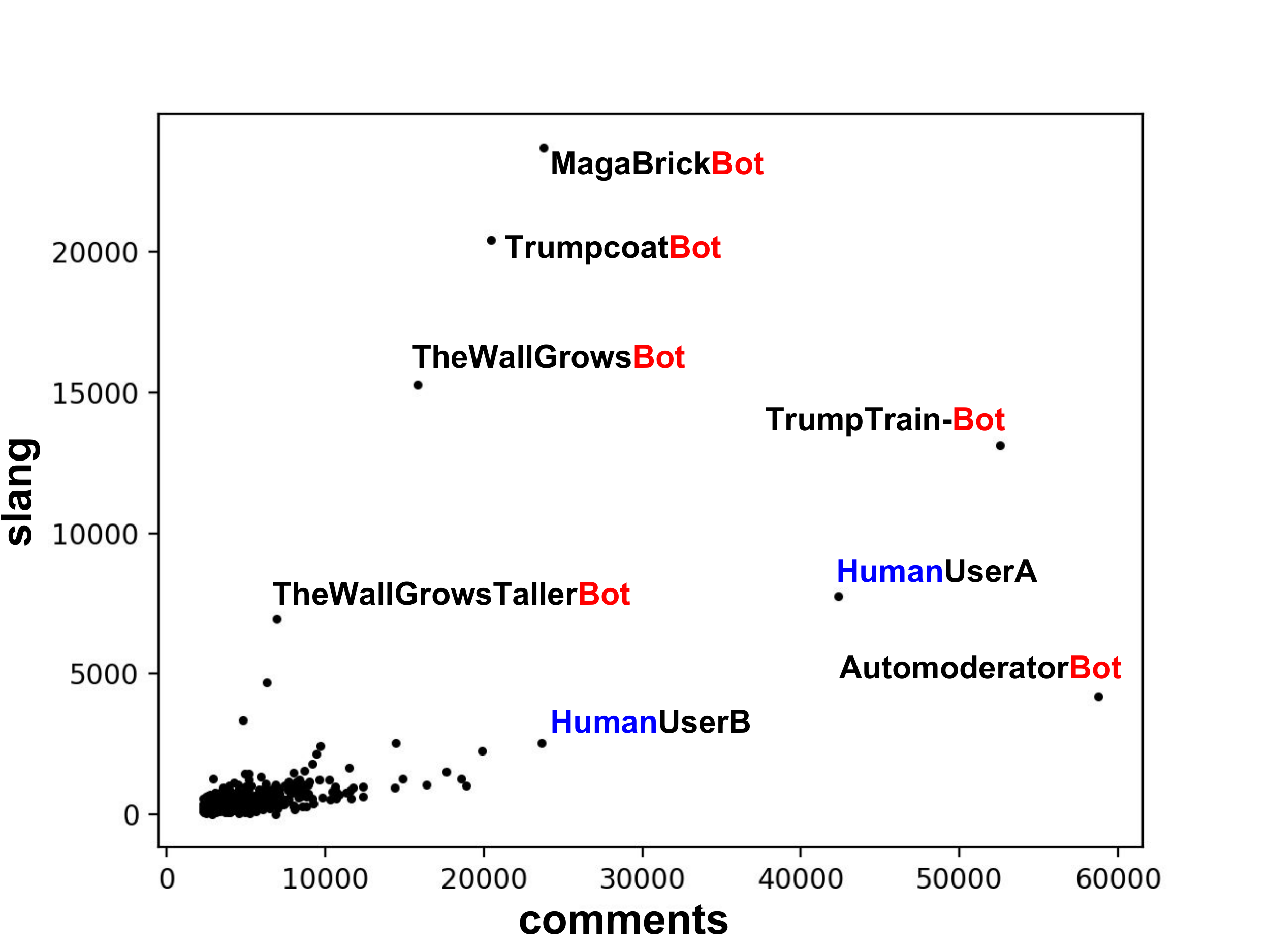}
  \caption{Slang used in Comments}~\label{fig:cslang}
\end{figure}
\subsection{Results} 
The most active accounts on \texttt{T\_D} are a mix of bots and humans. 1\% of these  accounts had the term ``bot'' in their username, e.g., ``TrumpTrainBot''. We considered that these were potentially bots or at least humans pretending to be bots. Their top 20 words with highest TF-IDF scores (for both potential bots and humans) corresponded to pro-Trump slang, such as: ``MAGA'' or ``Centipedes''. We also observed that the humans in this group constantly shared the same YouTube videos, usually whimsical videos promoting Trump. For instance, the following video \url{youtu.be/hgM2xN5TPgw}, a comical song about the ``Trump train'' (term used to refer to the movement of American voters supporting Donald Trump) was mentioned 54,550 times by these individuals. 

Given the importance of slang among these most active users, we refined our analysis of their jargon. Figure~\ref{fig:cslang} visualizes the relationships between the most active users' number of comments (X axis) against the number of comments they generated with slang (Y axis). Some of the accounts using the most slang were bots (they appear higher in the Y axis). Although bots were not a majority among the most active users (only 1\% of all accounts), they produced the majority of the content with pro-Trump slang. However, these bots do appear to have been successful in engaging the community: the bots only generated content if people replied to them or if someone mentioned pro-Trump slang. If we see that the bots are among the most active it is because people were actively interacting with them or using their related slang words. 
The humans and the bots in \texttt{T\_D} constantly engaged with each other by commenting on each others' posts. Humans engaged with the bots to play games with them. The most active bots (Figure \ref{fig:cslang}) had a gamification component and were activated when participants used certain slang words or mentioned the bot. For example, in the ``TrumpTrainBot'' users are in charge of ``moving forward'' the ``Trump Train.'' This bot increases its speed each time people use pro-Trump slang or reply to the bot. This bot generated a total of 54,540 comments. 
An example message from the ``TrumpTrainBot'' included: 
\begin{displayquote}``WE JUST CAN'T STOP WINNING, FOLKS THE TRUMP TRAIN JUST GOT 10 BILLION MPH FASTER CURRENT SPEED 175,219,385,117,000 MPH. At that rate, it would take approximately 9.209 years to travel to the Andromeda Galaxy (2.5 million light-years)!''.
The bot ``Trumpcoatbot'' was activated 21,569 times, while the bots ``MAGABrickBot'', ``TheWallGrowsBot'' and ``TheWallGrowsTallerBot'' which referenced ``building a wall'' were activated 27,187, 23,373 and 7,609 times respectively, and had a similar game dynamic as the ``TrumpTrainBot''.
\end{displayquote}
These gaming mechanisms might have helped to entertain and keep \texttt{T\_D} participants active.

Most of the top words used by these highly active participants were slang words, but with a pejorative connotation, such as ``cuck''. While the word is a derogatory slang term for a weak, or inadequate man, the word gained popularity in T\_D to denote conservatives (``cuckservatives'' ) who lacked ``real leadership'' and challenged Donald Trump on his spelling, his logic, or his facts \cite{WhyAngry79:online}. The word was used 90,047 times. In contrast, mainstream profanity like ``fuck'' was used 252,754 times. The use of specialized slang and profanity within the community was highly popularized.

\section{RQ2: Styles for Calling a Political Troll Community to Action}
Research Question 2 asked, ``How does a political trolling community mobilize its members to action?'' What ``styles'' were adopted to mobilize the community? (i.e., ways of calling the community to action). Do different call to action styles exhibit different engagement patterns?  Are they using offensive messages?

\subsubsection{Identifying Call to Action Styles.} To uncover how individuals tried to mobilize \texttt{T\_D}, we first identified all posts that were potential calls to action. We used lists of action verbs to flag posts with possible calls to action.  Through this process we identified 5,603 posts. Next, we hired three English-speaking college educated crowd workers from Upwork to categorize whether each of the posts made a call to action or not (it could be that the post had action verbs, but did not try to mobilize others.) The two coders agreed on 3,274 posts (Cohen's kappa: 0.58: Moderate agreement). We then asked a third coder to label the remaining posts upon which the first two coders disagreed. We used a ``majority rule'' approach to determine the category for those remaining posts. 

Once we had the posts categorized into ``call to action'' and ``non call to action'' posts, we identified the authors of the posts and characterized them based on their ``style'' for making calls to action. We considered each individual had a particular style for mobilizing others. Our goal was to cluster together the individuals that used the same style. For this purpose we represented each individual, as a vector where the components of the vector were: (1) the likelihood of making a call to action with a link attached: number of calls to action the person made with a link over their total number of calls to action; (2) likelihood of mentioning a public figure or organization: number of times public figures or organizations were mentioned over the total number of words in her calls to action (notice that similar procedure was followed for slang and profanity); (3) likelihood of using slang; and (4) likelihood of using profanity, (5) the median number of words in the call to action over the median number of words the person used in general in her posts.

After this step, each user was represented as a feature vector of size 5. Notice that the features we considered were similar to those used by prior work to characterize the messages of deviant groups \cite{savage2015participatory}. We then used a clustering method to group similar feature vectors (people with similar call to action styles). We use mean shift algorithm \cite{cheng1995mean} to group together similar vectors because it provides a non-parametric density estimation, and therefore does not require a prior for the number of clusters beforehand (unlike K-means). Mean shift helps us to discover different clusters of people, where each cluster represents groups of individuals with a particular style for making calls to action. Next, we inspected each cluster in detail to better understand the call to action style being used. In particular, we looked at: the number of calls to action and words used; distinct words present in the calls to action; public figures and organization mentioned; slang and profanity used.

\subsection{Results.}
\begin{figure}
\centering
  \includegraphics[width=1.0\columnwidth]{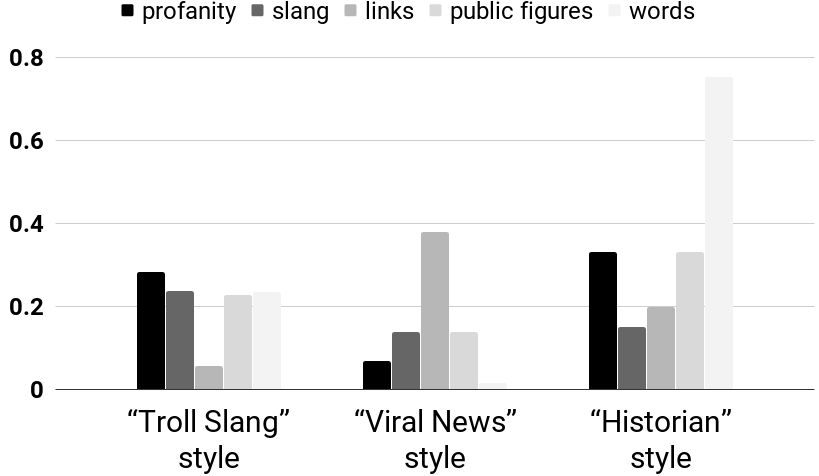}
  \caption{Features in each Call to Action Style. The Troll Slang style used the most slang in its calls to action; the Viral news style referenced the most links; and the Historian style used the most number of words intermixed with profanity in its calls to action.}~\label{fig:feature}
\end{figure}

1,608 of the posts generated by the most active \texttt{T\_D} participants were calls to action, and 1,350 individuals made those calls to action (almost 40\% of the most active individuals on \texttt{T\_D} called the community to action at least once.) These individuals had three main styles for making those calls to action (i.e., we discovered three main clusters). Table 1 presents an overview of the different styles and the community's engagement. Figure~\ref{fig:feature} shows the clustering results that distinguish the three call to action styles. Within each cluster, the histograms indicate the proportion of each of the features. Notice that we follow an  approach that is similar to that of prior work to visualize the clusters \cite{hu2014we}.

\begin{table}
\centering
\label{table:callstoaction}
\begin{tabular}{|l|l|}
\hline 
\multicolumn{1}{|c|}{Call to Action}              & \multicolumn{1}{c|}{Engagement Stats}                                                                                    \\ \hline
{\bf ``Historian'' style}                                   & \begin{tabular}[c]{@{}l@{}}{\bf \small  Number of Upvotes:}\\ Min = 83, Max = 30,226, \\ $\mu = 2,890.32$, $\sigma = 3,808.81$\\ {\bf \small Number of Comments:}\\ Min = 1 Max = 2069,\\ $\mu = 92.69$, $\sigma = 188.48$\end{tabular} \\ \hline
{\bf ``Viral News'' style}                                 & \begin{tabular}[c]{@{}l@{}}{\bf \small Number of Upvotes:}\\ Min = 84, Max = 15,625, \\$\mu = 2,404.85$, $\sigma = 2,497.64$ \\ {\bf  \small Number of Comments:}\\ Min = 1 Max = 511,\\ $\mu = 84.73$, $\sigma = 138.37$ \end{tabular} \\ \hline
\multicolumn{1}{|c|}{{\bf ``Troll Slang'' style}} & \begin{tabular}[c]{@{}l@{}}{\bf \small Number of Upvotes:}\\ Min = 81, Max = 32,970, \\ $\mu$  = 2,330.01, $\sigma = 3,735.77$\\{\bf \small Number of Comments:}\\ Min = 1, Max = 1907, \\ $\mu = 55.60$, $\sigma = 127.28$ \end{tabular} 
\\ \hline\end{tabular}
\caption{Descriptive Statistics Calls to Action Styles}
\end{table} 

{\bf Cluster 1: ``Troll Slang style''}: The individuals using this style represented 40\% of those who made calls to action.  The individuals in this cluster used the most ``pro-Trump slang'' (Figure \ref{fig:feature}). For instance, the following call to action uses the slang: {``deplorables'' and ``centipedes,''}: 
\begin{displayquote} 
``Calling SPANISH-SPEAKING DEPLORABLES!!! Want to do something that will make a  difference and win Trump VOTES? [...] Share with any Latino Centipedes!''
\end{displayquote}

The slang word most used was \emph{``kek''}. That word alone was present in over 14\% of the calls to action of this group. Notice that ``Kek' represents an Egyptian god that is depicted with a frog head and thus holding similarities to the meme of ``Pepe the frog,'' linked to the pro-Trump and alt-Right movements. Given this distinct usage for specialized words and icons, we refer to this call to action style as the ``Troll Slang'' style. 

One popular call to action of this cluster revolved around ``deporting'' people. The term ``deportation'' was used in almost 15\% of their calls to action. This word was used in two contexts: the first was regarding the deportation of immigrants from the US. The second was related to a ``troll button'' that participants could use against others \cite{Afewword98:online}, to call out trolling (or undesirable) behavior within the community:

\begin{displayquote}
``Time to hit that deport button on these trolls today...don't be scared...just do it''. 
\end{displayquote}

This style also had the particularity that it called people to action towards causes that Trump's opposition  labeled as ``conspiracy theories'' or ``fake news stories''. For instance, certain calls  requested people to take action in response to conspiracies surrounding the murder of Democratic National Committee employee Seth Rich \cite{TheSethR95:online}.  An example: 
\begin{displayquote}
``Computer Wizards and Artists: Can we please get a good infographic on the Seth Rich murder with known facts and inconsistencies to share on social media? Don't let cucked admins derail us now, this case is cracking at the seams!''
\end{displayquote}

This style was also the one that used profanity the second most. The profanity was usually used against Trump's opponents. In general, the style of these calls to action appear to be based on creating a collective identity by utilizing slang supporting their political believes and promoting causes or stories that their opposition deems to be unreal. The creation of such collective identity might facilitate driving people to action, as language and the use of slang solidifies the identity of a group \cite{sarabia2004maintaining}. However, this style was in general the one that received the lowest engagement from the community (this style received the smallest median number of comments and upvotes). The slang and ``conspiracy theories'' might have been difficult for newcomers or less committed community members to follow, and hence their participation was more limited. However, we did observe that this call to action style was able to generate the largest number of upvotes for one post that called the community to take action and report an NBC employee that mocked Trump's 10-year old child, Barron: 

\begin{displayquote}
``SNL writer calls Barron Trump a homeschool shooter be a shame if a bunch of us forwarded this to NBC''
\end{displayquote}

{\bf Cluster 2: ``Viral News'' style}: The people in this cluster (9\% of all the individuals making calls to action) had a style that focused primarily on posting a link to a news story and then asking people to share and popularize on different social media platforms. This cluster shared the most number of links in its calls to action (Figure~\ref{fig:feature}). Given this behavior of distributing news, we named this style ``Viral News''. 

An example call to action: 
\begin{displayquote}
``It has arrived. The Top Hillary Crimes in one list. FBI docs, Wikileaks, and more. GET IN HERE! [...] This list needs to be at the top every single day until the election. MAGA [...] Share this short link on Twitter and Facebook with the usual hashtags http://redd.it/59sh7p''
\end{displayquote}

This call to action style also used the least slang and practically did not use also any profanity, only 6\% of its calls to action had profanity and it used the least amount of words (Figure~\ref{fig:feature}). While this style had the least number of people involved, and the calls to action were the most sporadic, the style in general ensured that large crowds participated in its calls to action (its average number of upvotes and comments were more than the ``Troll Slang'' style). It might have been that the straightforward aspect of the call to action facilitated mobilization (even despite its inconsistency and limited number of organizers).

{\bf Cluster 3: ``Historian'' style}: The calls to action in this cluster were generated by 51\% of the individuals who were making the calls to action. This style focused on first explaining in detail the political ecosystem and then requesting people to take a specific action. This style was the one with the most amount of words in proportion with the others (Fig. ~\ref{fig:feature}). This dynamic of explaining the background information to actions being requested lead us to name this style the ``Historian.'' An example: 

\begin{displayquote}
``Tomorrow the House is scheduled to vote to eliminate consumer privacy online. Call your representatives. Tell them to vote NO [...]  net neutrality and your privacy online is something that every side must fight for. Particularly if you're a conservative who doesn't want to have everything they do be sold to the highest bidder [...] The founding fathers of this country believed in certain unalienable rights [...]  If the founding fathers were alive today, protecting your privacy online would be paramount on their list.''
\end{displayquote}

This style mentioned the most public figures and the most profanity in its calls to action  (32\% of the calls to action referenced public figures or organizations and 33\% of the calls to action used profanity. The profanity used was against political situations and was rarely used to attack political opponents). This style did not directly give people orders, but rather suggested which actions were needed. Suggestions to take action were also frequently followed by the assertion that said action was already performed by others, including the person that made the suggestion in the first place. For instance, the following post explains to people why they should eliminate their Netflix account (as it supports the opposition's ideology), and showcases that the poster has herself already cancelled her account:
\begin{displayquote}
``...All the more reason that we should dump Netflix. Their "Dear White People Show" they released is a bit too preachy.  If Soros is funding, hell yes, I'm done. Pic related:  My cancellation receipt..''
\end{displayquote}
This approach might be helpful to convince people to take action in their effort to follow the example of their peers \cite{banerjee1992simple,milgram1969note}. The ``Historian'' style was the most effective call to action style in terms of the average number of comments and upvotes it secured (it was the call to action style that was able to more consistently assure action from a large crowd in \texttt{T\_D}).

\subsubsection{Understanding Engagement and Call to Action Style.}
\begin{figure}
\centering
  \includegraphics[width=1.0\columnwidth]{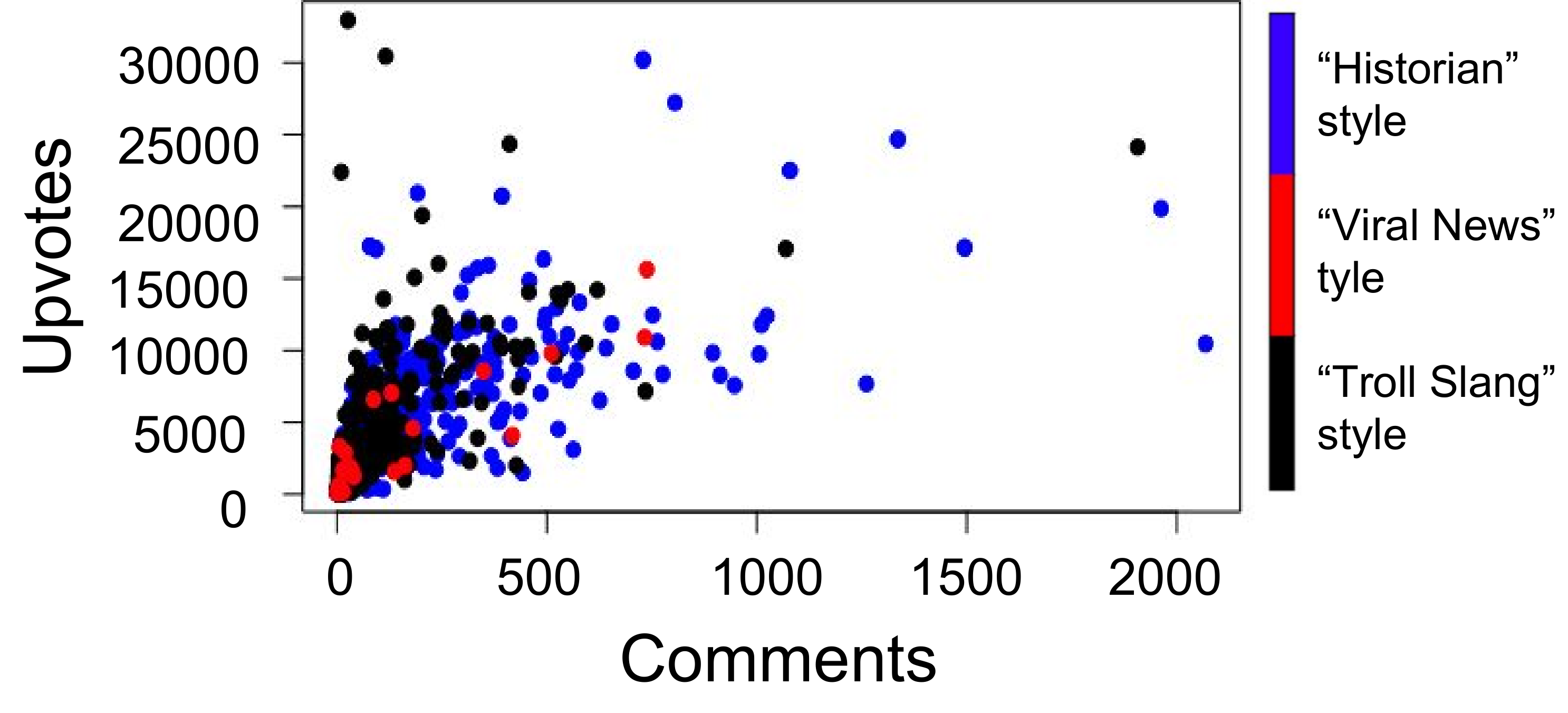}
  \caption{Number of upvotes and comments that each call to action received. We color code call to actions based on their style.}~\label{fig:clustersposts}
\end{figure}

To understand how the community reacted to each call to action style, we plotted the number of upvotes (X axis) and number of comments (Y axis) that each call to action received and color-coded them according to the style it belonged to. Figure \ref{fig:clustersposts} shows that the ``Historian'' style appears to have continuously obtained a high number of upvotes and comments. However, the ``Troll Slang'' style occasionally managed to secure more engagement than the ``Historian'' style. But this was not common. The ``Viral News'' style although it was more rarely used, received similar number of upvotes and comments regardless. The community usually engaged with this style less than with the ``Historian'' style. But it received occasionally more engagement than the ``Troll Slang'' style.

\begin{figure}
\centering
  \includegraphics[width=1.0\columnwidth]{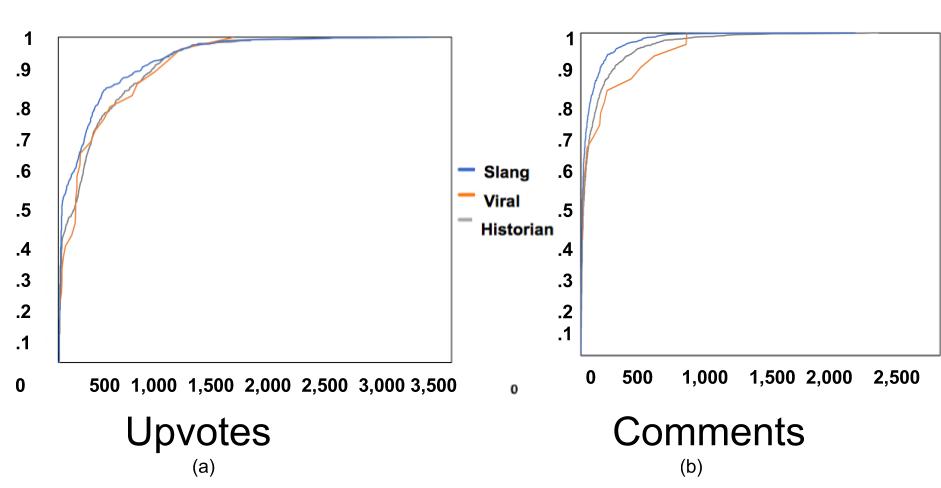}
  \caption{Cumulative distribution associated with (a) Upvote rate (b) Comment rate  and}~\label{fig:cdf}
\end{figure}

To further understand if a call to action style exhibited a different engagement pattern than other styles, we computed the number of upvotes and comment that each of its related posts received and plotted the corresponding Cumulative Distribution Functions (CDF) as seen in Figure \ref{fig:cdf}. To determine whether there are significant differences between the three datasets (i.e., between the number of comments and upvotes that a type of call to action style  receives), we used a Kruskal-Wallis H test to investigate the similarities/differences between the number of comments and upvotes across call to action styles. The test statistic for the number of upvotes was
H = 23.066 with p-value \textless .001.  
The test statistics for the comments was H = 33.587 with p-value \textless .001. Therefore, the median number of comments that each style received was significantly different.  Overall, the community's engagement with each call to action style was different.  We believe that the ``Historian'' style helped to maintain the community continuously engaged and on the same page about what was occurring in the political ecosystem. But the ``Viral News'' style helped to rapidly mobilize the community at a given point in time (that was why these calls to action could be sporadic and without adopting any of the community slang).

\section{Discussion}
We discuss our results using the previously described themes that facilitate collective action: framing processes, mobilizing structures, and opportunity structures. We finish by also discussing implications of our research. 

\subsection{Mobilization Structures.} \texttt{T\_D} presented three distinct styles or strategies for mobilizing the community. The ``Historian'' style was the most effective for retaining and mobilizing a large crowd of highly-active community members. Civic technology research argues that establishing opponents can be mobilized by narrating the ideals of what is being fought for \cite{dimond2013hollaback}. In this case, the ``Historian'' style facilitated long term participation because its detailed explanations helped individuals to understand the community's reasoning in the current political ecosystem and thus take action. Therefore, adequately \emph{framing} what is occurring is important in mobilization. 

\subsection{Framing Process.} The ``Troll Slang'' style integrated pro-Trump vocabulary into its calls to action, likely helping to frame and narrate the collective identity of \texttt{T\_D}. Previous work in online communities has identified the importance of building an identity with community members in order to facilitate mobilization \cite{kraut2012building}. Notice, however, that in online groups opposing the establishment if the goal is to mobilize, it might be better to focus on sharing the community's cause rather than the vocabulary that the community uses.  

The ``Troll Slang'' style discussed ``conspiracy theories'' \cite{conspiracy2017} to a significant extent. The term ``conspiracy theorist'' originated in the 70s with the CIA as an effort to refute anyone who challenged official narratives. Therefore, when we see trolls engaging with alleged conspiracy theories, it means they are framing events in a way that their opposition (in this case the ``Establishment'') has somehow discredited. Engaging in conspiracy theories thus also helps to promote \texttt{T\_D}'s  collective identity  as it establishes a dynamic where \texttt{T\_D} promotes stories that the community believes are true versus the stories that the others (the opposition) discredits. 

\subsection{Opportunity Structures.}  Calls to action following the ``Viral News'' style had a very straightforward way of requesting action and was the most effective at mobilizing the largest crowds. This resembles the findings of research in online civic platforms where direct requests usually garnered more participation than ``more manipulative'' calls to action  \cite{savage2016botivist}.  In this case, being direct and simple might have been a better opportunity to more easily mobilize a much larger mass. Individuals did not have to invest much in reading about the problem being addressed, or spend time learning the culture. They simply focused on executing. The ability to rapidly take action is an opportunity structure that facilitated participation in this case. This is something that the ``Troll Slang'' style was never able to accomplish with posts that called ``all centipedes'' into action. However the ``Historian'' style might be helping to indoctrinate newcomers about the community's belief system and the reasons behind it, which in turn facilitated afterwards the mobilization of participants at the time it was needed without much explanation about the reasons.
One characteristic of the ``Historian'' style is that it occurred more frequently. The ``Historian'' style was adopted to promote the community's belief system, but this also created opportunities to drive the community to take rapid action when needed without having to explain much in detail (``Viral News'' style).

Bots within reddit have been used mainly for automated support for human moderators \cite{long2017could}, upvoting or downvoting posts \cite{gilbert2013widespread} and for entertainment \cite{massanari2016contested}. However, \texttt{T\_D} subreddit to our knowledge was one of the first to opportunistically use bots for creating an identity and engagement in a political context.

It has been found that design structures from games, such as levels, achievements, points, and leader boards increase people's motivation to participate in an online community \cite{iacovides2013games}.  \texttt{T\_D} might be using bots to opportunistically drive more participation through games. Notice also that bots such as ``TrumpTrainBot'' or ``TheWallGrowsBot'' used slang which  likely also provided the opportunity to promote a sense of identity within \texttt{T\_D} \cite{short2006studying,sarabia2004maintaining}.

\subsection{Implications for Collective Action Theories}
Research in collective action sustains that a successful way to mobilize people is by effectively communicating the goals and purpose of the movement \cite{bennett2011digital}, as well as developing a sense of identity among participants  \cite{kraut2012building}. Our findings align with these theories as the calls to action in \texttt{The\_Donald} that were able to retain and mobilize the largest number of participants, were the ones that explained in detail the motives of why they were calling to action and what they were trying to accomplish. What is interesting is that our research uncovered how within subversive environments, where the community had large opposition, it was more important to explain the motives of the organization instead of promoting an identity for the community. This was essential in order to rapidly mobilize more people.

\subsection{Implications for Designers of Civic Media}

Our results suggest that taking the time to explain the political ecosystem and showing that they already have taken some sort of action could have been crucial in making people on \texttt{T\_D} participate in collective action long term in their effort to follow the example of their peers. Designers of civic technologies could take this into account to create online tools that help politicians and governments to more easily trigger their long term participation in particular political endeavors. Social networks such as Facebook and reddit exist, however they are not tailored to coordinate collective action. We believe there is value in interfaces that can inform governments of the best ways to call citizens to action given what has been organically observed to work online. Such tools could inform governments of how to frame their calls to action to effectively mobilize their supporters. In the future we will explore interfaces that for a wide range of online political communities can automatically detect how calls to action are made for a variety of causes, and for a given goal can inform best ways to mobilize citizens.

Some of the most interactive experiences on \texttt{T\_D} centered on engagement with bots, especially by playing games with them. Future work could investigate the effectiveness of this type of gaming mechanisms to promote culture or ideology in a community. This seems to be an effective way for this trolling community to engage its members. Designers might also consider how to make the promotion of a community's culture a game. For example, one could imagine creating leadership boards for individuals who were able to mobilize newcomers to adopt the community's slang. Given also that people built bots that were specifically tailored for \texttt{T\_D}, we can imagine there is also value in facilitating tools through which citizens can rapidly understand what technology has been created to promote their ideology, and enabling them to rapidly be able to build off and improve it, maybe even have the development of tools be part of a game. The novel aspect of  our design proposition is to incorporate community culture promotion and gamification as a core design factor to civic media, for facilitating ideology or collective identity building. 

\subsection{Limitations and Future Work} 
More empirical work needs to be carried out to understand the connections between troll's individual feelings of belonging, commitment and identification to a community, and how much this results in the community's  mobilization, and how the community's collective identity is showcased. Future work could also focus on understanding the processes involved in creating ``strong'' or ``weak'' collective identities in communities of online trolls. Another direction of future work is understanding how trolls' collective identity inter-plays in whether people drop out of the movement that the troll community is promoting.  

The insights this work provides are limited by the methodology and population we studied. For example, the subreddit we examined, \texttt{T\_D}, focuses around US politics, especially the political campaign and presidency surrounding Donald Trump. Hence our results might not describe how other similar communities behave. 

Our investigation also focused on breath, rather than on depth. As a result, we do not know much about the identities and motivations of the people participating in \texttt{T\_D}. Future research would be wise to conduct detailed interviews with the participants of this subreddit.

\section{Conclusion}

In this article we investigated the subreddit \texttt{T\_D} as a vehicle for understanding sustained participated and collective action production within political troll communities. In \texttt{T\_D} to mobilize others it was most important to explain in detail the political ecosystem and educate the public about the meaning of certain events. This was the most effective strategy in mobilizing other \texttt{T\_D} participants long term (highest retention). The individuals with most sustained participation used socio-technical tools, such as bots, to maintain themselves engaged and entertained. Our findings can help to design novel civic media and troll moderation strategies.

\bibliographystyle{aaai}
\bibliography{references}

\end{document}